\documentclass[twocolumn,showpacs,preprintnumbers,amsmath,amssymb]{revtex4}
\usepackage{amsmath}
\usepackage{graphicx}
\usepackage{dcolumn}
\usepackage{bm}
\usepackage{multirow}
\usepackage{color}
\usepackage{MnSymbol}

\begin{document}


\title{Experimental investigation of pair dispersion with small initial separation in convective turbulent flows}

\author{Rui Ni and Ke-Qing Xia}
\address{Department of Physics, The Chinese University of Hong Kong, Shatin, Hong Kong, China}

\author{}
\affiliation{}


\date{\today}

\begin{abstract}
We report an experimental investigation of pair dispersions in turbulent thermal convection with initial separation $r_0$ ranging from sub-Kolmogorov scale to scales in the inertial range. In the dissipative range of scales we observed for the first time in experiment the exponential growth of the separation between a pair of particles predicted by Batchelor and obtained a Batchelor constant $0.23\pm0.07$. For large $r_0$, it is found that, for almost all time range, both the mean-square separation and distance neighbor function exhibit the forms predicted by Batchelor, whereas the two quantities agree with Richardson's predictions for small $r_0$. Moreover, the measured value of the Richardson constant $g=0.10\pm0.07$, which is smaller than those found in other turbulence systems. We also demonstrate the crossover of the mean-square separation from the exponential to the Batchelor regimes in both temporal and spatial scales.

\end{abstract}


\maketitle


Turbulent relative dispersion of a pair of particles is of central importance to a wide range of natural processes such as pollutant spreading in the atmosphere \cite{2001JASHuber} and mixing in oceans. The concept was first introduced by Richardson, who attempted to explain the large observed value of turbulent diffusivity in the atmosphere. He introduced a quantity named distance neighbor function (DNF) and the diffusion equation to describe the evolution of DNF \cite{1926PRSLARichardson}. With Kolmogorov's scaling theory, Obukhov refined Richardson's prediction and found that $\langle r^2\rangle=g\epsilon t^3$, with $r$ the pair separation, $\epsilon$ the mean kinetic energy dissipation rate, and $g$ is a dimensionless constant called the Richardson constant \cite{1941Obukhov}. Batchelor \cite{1950QJRMSBatchelor}, recognizing that over short time the initial separation $r_0$ between the pair of particles would be important, obtained $\langle |{\bf r}(t)-{\bf r}_0|^2\rangle =f(r_0)t^2$ for $\tau_{\eta} \ll t \ll t_0$, where $\tau_{\eta}=(\nu/\epsilon)^{1/2}$ is the Kolmogorov time scale and $t_0 = (r_0^2/\epsilon)^{1/3}$ is a characteristic time below which the initial separation is important. The Richardson-Obukhov scaling is now supposed to hold for $t_{0} \ll t \ll t_L$, where $t_L$ is the integral time scale. In the above $f(r_0)= [D_{LL}(r_{0})+2D_{NN}(r_{0})]$, with $D_{LL}$ and $D_{NN}$ being the 2nd-order longitudinal and transverse Eularian structure functions (ESF) respectively.  

An important regime of pair dispersion is the very early stage of separation in which relevant spatial scales are within the dissipative subrange \cite{2001ARFMSawford, 2009ARFMSalazar}. Turbulent dispersion and mixing in this regime is closely related to the reaction rate for fast reacting scalars, such as in combustions.  Batchelor was the first to argue that the growth rate of pair separation should be proportional to the separation distance $r$ itself in this regime  \cite{1952PRSLABatchelor}, which leads to an exponential growth
 $\langle r^2\rangle\sim r_0^2exp(\xi t)$ when both the initial and final particle separations are within the dissipative range, $r_0^2\ll\langle r^2(t)\rangle\ll\eta^2$. Here the growth rate  $\xi=2B/\tau_{\eta}$ and $B$ is called the Batchelor constant. However, in most previous experimental studies, the initial separation is in the inertial subrange \cite{2006NJPOuellette, 2000JFMOtt}. As a result, the exponential growth regime has never been observed in experiments and the Batchelor constant has never been measured. 
 
Recently, we have shown that the Lagrangian particle tracking velocimetry (PTV) can be applied to thermally-driven turbulent flows and have obtained particle pairs with separations smaller than $\eta$ \cite{2012JFMNi}. One advantage of our system is that the range of its parameters is such that both the dissipative and inertial subranges can be easily accessed in the experiment. In fact, we have accurately determined the energy dissipation rate $\epsilon$ from the measured dissipative range ESFs using PTV \cite{2011PRLNi}. In this respect, turbulent thermal convection provides a good platform for studying properties of particle dispersions in both the dissipative and inertial subrange in a single experiment. From a practical point of view, studying two-particle dispersion in turbulent thermal convection is important in understanding the motion of passive scalars in the atmosphere and oceans in which buoyancy is a relevant driving force and is absent in most previous studies. 

In this Letter, we report measurements of the mean-square separation of a pair of particles  $\langle {\bf r}^2(t)\rangle$ (and also $\langle |{\bf r}(t)-{\bf r}_0|^2\rangle$) in buoyancy-driven turbulent thermal convection. Lagrangian particle tracking velocimetry (PTV) was used in the experiments, which were carried out in a cylindrical cell with water as working fluid. The measurements were made in the cell center with $r_0$ ranging from dissipative to inertial range of scales. 
The height and diameter of the cell both equal to 19.2 cm, so the aspect ratio is one. The experiments were conducted at fixed Prandtl number $Pr=\nu/\kappa$=4.4 with various Rayleigh number $Ra=\alpha g\Delta TH^3/\nu\kappa$ (from $2.9\times10^9$ to $1.3\times10^{10}$), here $g$ is the gravitational acceleration, $\Delta T$ the temperature difference across the fluid layer, and $\alpha$, $\nu$ and $\kappa$, respectively, the thermal expansion coefficient, kinematic viscosity and the thermal diffusivity of the working fluid. To compare the results with other turbulence systems, the micro-scale Reynolds numbers $R_{\lambda}$ are determined by using $R_{\lambda}=\sqrt{15u'^4/\epsilon\nu}$ with $u'$ the root mean square velocity and $\epsilon$ the energy dissipation rate in the cell center \cite{2012JFMNi}.  The tracking volume [$\delta V\simeq$(5 cm)$^3$] in the center of the cell was illuminated by a laser beam, and the scattered light from the seeding particles (diameter $d_p=50$ $\mu$m polyamid, density $\rho=1.03$ g/cm$^3$) were captured by three cameras simultaneously. The Stokes number $St=\tau_p/\tau_{\eta}$ ranges from $10^{-4}$ to $10^{-3}$, with $\tau_p$ being the time scale of the Stokes viscous drag due to interaction between particle and fluid. The number is much less than one, indicating that the particles would be safely regarded as tracers. As Kolmogorov time scale is 0.3$\sim$0.5 s in the parameter range of our experiment and the camera frame rate is 50 or 100 fps depending on the $Ra$, the temporal resolution is sufficient to resolve dissipative range properties. The error of the particle position after calibration is $\sim$ 8 $\mu$m. However, due to the finite particle size and the diffraction effect, the minimum resolvable separation between a pair of particles is typically between 100 $\mu$m to 200 $\mu$m, which is less than $\eta\simeq0.5$ mm in the experiment. Thus we are able to determine pair separations with initial separation $r_0$ smaller than the Kolmogorov length scale. In practice, we binned all pairs of particles with initial separation in the range $[r_0-\delta r_0,r_0+\delta r_0]$ when computing statistics. For $r_0<20\eta$, we take $\delta r_0\approx\eta/5$ and for $r_0\geq20\eta$, $\delta r_0\approx\eta$. 
Other details of the setup and calibration have been described elsewhere \cite{2012JFMNi}. 

\begin{figure}[htbp]
\begin{center}
$\begin{array}{cc}
\includegraphics[width=3.4in]{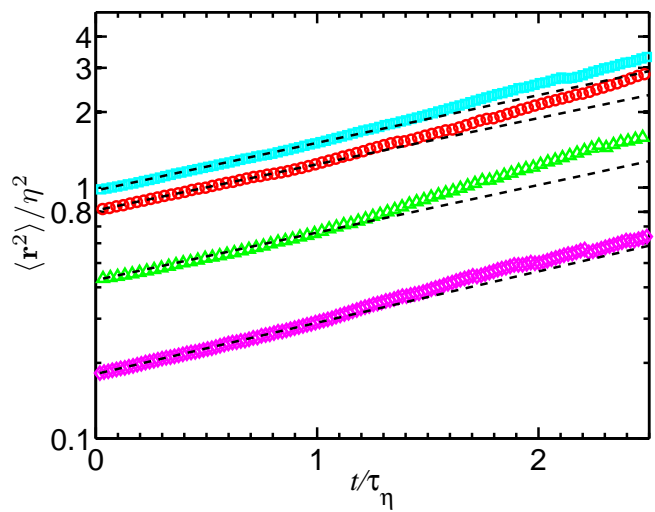} 
\end{array}$
\caption{(Color online) The mean square separation $\langle {\bf r}^2\rangle$ as a function of time for four different Rayleigh numbers with initial separation smaller or close to $\eta$ in a semi-log plot. From top to bottom,  cyan squares ($Ra=1.0\times10^{10}$, $R_{\lambda}=67$, $r_0=\eta$); red circles ($Ra=1.3\times10^{10}$, $R_{\lambda}=84$, $r_0=0.9\eta$); green triangles ($Ra=6.1\times10^{9}$, $R_{\lambda}=53$, $r_0=0.7\eta$); pink diamonds ($Ra=2.9\times10^{9}$, $R_{\lambda}=35$, $r_0=0.4\eta$). The dashed lines show the exponential fit to the respective data for $t\le\tau_{\eta}$.}
\label{fig:diss_exp}
\end{center}
\end{figure}

Figure~{\ref{fig:diss_exp}} plots in semi-log scale four mean square separations all with $r_0$ within the dissipative scale.  It is seen that for $t<\tau_{\eta}$ and $r<1.5\eta$ all curves with different Ra ($R_{\lambda}$) do follow an exponential growth as shown by the respective dashed lines.  This initial exponential growth is observed for all nine values of Ra measured. From exponential fits we find no systematic Rayleigh (Reynolds) number dependency for the Batchelor constant, which leads to an average Batchelor constant $B=0.23\pm0.07$.  This number was first estimated by Batchelor and Townsend based on the assumption that the dissipative separation is mainly due to stretching from the velocity gradient, which gives a range $B= 0.35 \sim 0.41$ \cite{1956Batchelor}. However, it was later argued that the original estimate is too large due to the lack of persistence of the rate-of-strain tensor and the role of vorticity \cite{1990JFMGirimaji} and it only serves as an upper limit. There are several simulation \cite{1990JFMGirimaji} and model \cite{2005JFMChun} studies that attempt to estimate this constant and they give $B=0.093\sim0.13$. It is seen that our value of Batchelor constant is larger than previous findings but still smaller than the upper limit proposed by Batchelor \& Townsend \cite{1956Batchelor}. To our knowledge, the present result is the first experimental confirmation of the exponential regime and the direct measurement of the Batchelor constant.

\begin{figure}[htbp]
\begin{center}
$\begin{array}{cc}
\includegraphics[width=3.4in]{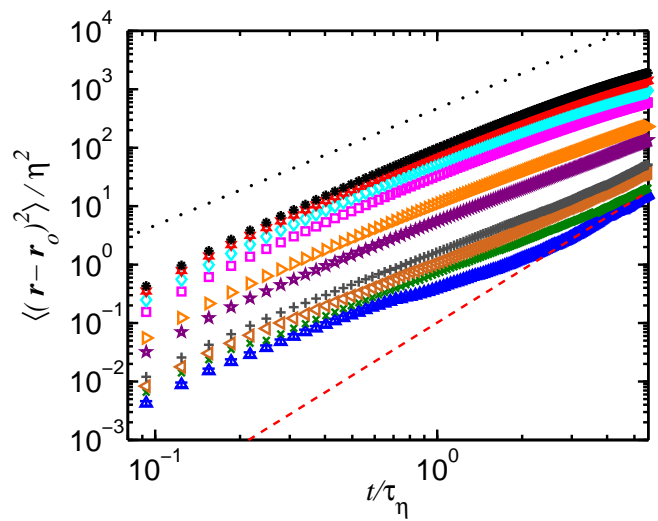} 
\end{array}$
\caption{(Color online) The normalized mean square separation $\langle ({\bf r}-{\bf r_0})^2\rangle/\eta^{2}$ as a function of time for different initial separations. From bottom to top, $r_0$=0.9$\eta$, 1.3$\eta$, 1.7$\eta$, 2.2$\eta$, 4.3$\eta$, 6.5$\eta$, 8.7$\eta$, 10.8$\eta$, 13.0$\eta$, 15.1$\eta$. The dotted line and dashed line indicate the Batchelor and Richardson regimes respectively.}
\label{fig:disp}
\end{center}
\end{figure}

\begin{figure}[htbp]
\begin{center}
$\begin{array}{cc}
\includegraphics[width=3.4in]{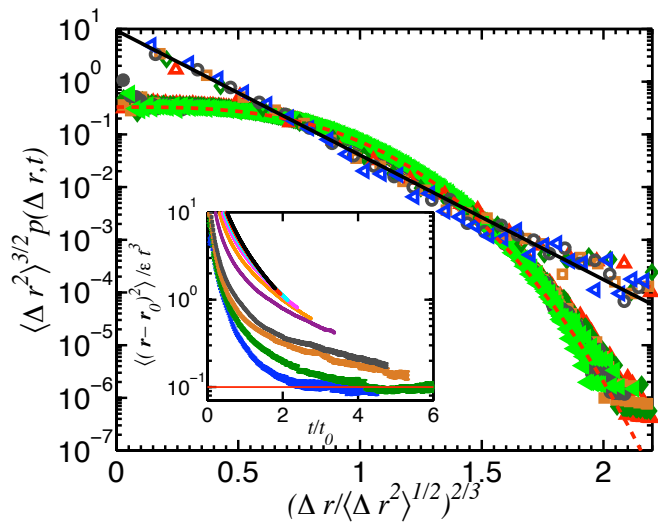} 
\end{array}$
\caption{(Color online) The distance neighbour functions for different initial separations at $Ra=1.3\times10^{10}$ ($R_{\lambda}=84$). The red curved dashed line is BatchelorÕs predicted PDF, while the black straight line is RichardsonÕs. The open symbols show the experimental results for initial separation $r_0=0.9\eta$ with time t ranging from $3.1\tau_{\eta}$ to $5.5\tau_{\eta}$. The closed symbols show the results for initial separation $r_0=52\eta$ with time t ranging from $5.5\tau_{\eta}$ to $8.1\tau_{\eta}$. Inset: The mean-square separation compensated by  $t^3$ with time normalized by $t_0$. The red solid line gives the Richardson constant $g=0.01$. From bottom to top, the curves represent different initial separations same as those in Fig.~{\ref{fig:disp}}. The thickness of each line at different times show the uncertainties for the data.}
\label{fig:PDF}
\end{center}
\end{figure}

Figure~{\ref{fig:disp}} shows the temporal evolution of $\langle |{\bf r}(t)-{\bf r}_0|^2\rangle$ for different $r_0$, which increases, from bottom to top, from dissipative range of scales to inertial range of scales. The dotted and dashed lines in the figure show Batchelor and Richardson scalings respectively. In PTV, the number of velocity pairs varies for different spatial separations, and pairs with the small separations generally have lower probability of being measured than those in some intermediate range of scales. Therefore, the one with smallest $r_0$ has lowest number of particle pairs for statistics. In our experiment, there are $10^7$ pairs of particles for $r_0=60.6\eta$, but only $10^3$ for $r_0=0.9\eta$. In Fig.~{\ref{fig:disp}} we show only the statistical errors for $r_0=0.9\eta$, as the uncertainty for this data is the biggest. One may note that even the largest error bar is within the symbol. For large $r_0$, there is a power law regime in a range of time scales extending from 0.1 to 6 $\tau_{\eta}$, whose exponent is very close to the one predicted by Batchelor \cite{1950QJRMSBatchelor}, i.e. $\langle |{\bf r}(t)-{\bf r}_0|^2\rangle\sim t^2$ (shown as the dotted line in the figure). For $r_0=0.9\eta$, there is no single power law for the entire range. For $t > 2\tau_{\eta}$ the behavior could be well described by the Richardson-Obukhov's law for two-particle diffusion, i.e. $\langle |{\bf r}(t)-{\bf r}_0|^2\rangle=g\epsilon t^3$. This can be seen more clearly in the inset of Fig.~{\ref{fig:PDF}} as the plateau for the square separation compensated by Richardson-Obukhov scaling.  Here it is seen that two data sets with the smallest initial separations ($r_0=0.9$ and $1.3\eta$) reach the plateau, whereas the others show similar trend towards the Richardson-Obukhov scaling but lack sufficient time to develop. Note that the time scale in the inset of Fig.~{\ref{fig:PDF}} is normalized by a characteristic time $t_0=(r_0^2/\epsilon)^{1/3}$, below which the initial separation is important. 
From the red solid line we obtain the Richardson constant $g= 0.10\pm0.007$. Previous studies show that there is a large uncertainty on the value of $g$. For non-buoyancy driven turbulent flows, more recent experimental and numerical studies \cite{2000JFMOtt, 2002PRLBoffetta,2004JFMYeung,2005POFBiferale, 2008POFSawford} suggest $g\approx0.5$. For thermal convection, a numerical study found that the value of $g$ equals to 0.16 and this smaller value was attributed to the correlated pair motion in thermal plumes \cite{2008PRLSchumacher}. As the plumes' motions are predominantly in the vertical direction, this would imply that pair dispersions should behave differently in different direction. However, by studying pair dispersion in the vertical and lateral directions separately we find the dispersion properties to be isotropic with the Richardson constant nearly the same in the vertical and the two lateral directions, i.e. each being 0.03. This suggests that pair dispersion in all directions are affected by some correlated motions. 
It is noted that flow in the cell center is affected by the large-scale circulation. This coherent motion is also azimuthally rotating, which may induce correlated motions in different directions.

The DNF represents the spherically averaged PDF for pairs of particles with separation r at time t, i.e. $p(r, t)$. Richardson first suggested that relative dispersion can be modeled by a diffusion equation for the DNF. For the isotropic flow, the diffusion equation can be expressed as $\partial p(r,t)/\partial t=(1/r^2)\partial[ r^2K(r,t)\partial p(r,t)/r]/\partial r$ with $K(r,t)$ being the diffusion constant. Richardson proposed that $K(r, t)=k_0\epsilon^{1/3}r^{4/3}$ based on the experimental measurements in the atmosphere, and found that $p_R(r, t)=\sqrt{143/2}\times429/70(\pi\langle r^2\rangle)^{-3/2}exp[-(1287r^2/8\langle r^2\rangle)^{1/3}]$. Assuming $K(r,t)\sim t^2$, Batchelor found another solution to the diffusion equation, i.e. $p_B(r, t)=(2\pi\langle r^2\rangle/3)^{-3/2}exp[-3r^2/2\langle r^2\rangle]$. The two solutions are shown in Fig.~{\ref{fig:PDF}} as black solid line (Richardson's prediction) and red dashed line (Batchelor's prediction).
In both solutions, the separation between two particles were assumed to be zero at the very beginning.  Experimentally, however, even one could resolve sub-Kolmogorov scale, the initial separation would be much larger than $0$. One way to solve this problem is subtracting all particle separations with their initial value $\Delta r=r-r_0$, and replacing $p(r,t)$ with $p(\Delta r,t)$ \cite{2006NJPOuellette}. In Fig.~{\ref{fig:PDF}}, the open symbols all have $r_0=0.9\eta$ and closed ones $r_{0}=52\eta$. There are five data sets at different times for each initial separation. The time are chosen to fall into the time range where the particle separations increase as $t^2$ ($r_0=52\eta$) and $t^3$ ($r_0=0.9\eta$) scalings respectively. It is clear that the DNF results agree with Richardson's prediction for small $r_0$ and agree with Batchelor's prediction for large $r_0$. 

\begin{figure}[htbp]
\begin{center}
$\begin{array}{cc}
\includegraphics[width=3.4in]{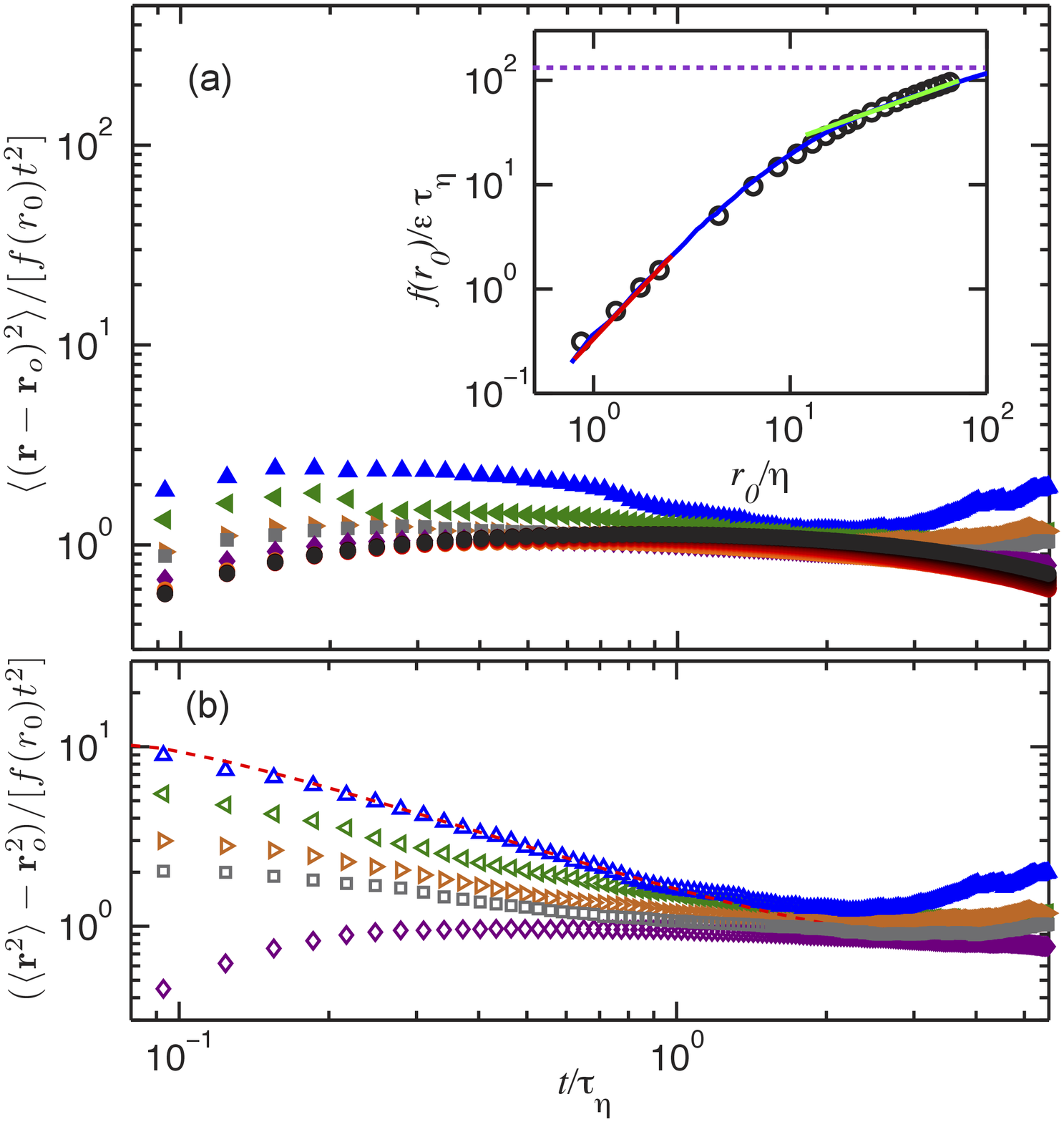} 
\end{array}$
\caption{(Color online) (a) The mean square separation $\langle({\bf r}-{\bf r}_0)^2\rangle$ compensated by $f(r_0)t^2$ as a function of normalized time $t/\tau_{\eta}$ for different initial separations measured at $Ra=1.3\times10^{10}$ ($R_{\lambda}=84$).  From top to bottom, $r_0$=0.9$\eta$ ($\filledmedtriangleup$), 1.3 $\eta$ ($\filledmedtriangleleft$), 1.7 $\eta$ ($\filledmedtriangleright$), 2.2$\eta$ ($\filledmedsquare$), 4.3$\eta$ ($\filleddiamond$). There are 18 data sets with $r_0=6.5\eta\sim65\eta$ that collapse onto each other, which are represented by the solid circles . 
Inset: The black circles represent the value of $\langle({\bf{r}}-{\bf{r}}_0)^2\rangle/t^2$ at $t=2\tau_{\eta}$ for different $r_0$ and the blue solid line represents $f(r_0)$ determined from Eulerian structure functions. Both quantities are normalized by $\epsilon \tau_{\eta}$. The red and green solid lines are $f(r_0)= \frac{1}{3}r_0^2/\eta^2$ and $f(r_0) = \frac{11}{3}Cr_0^{2/3}/\eta^{2/3}$ ($C=1.56$ is the Kolmogorov constant \cite{2012JFMNi}.), which are the dissipative and inertial range scaling predictions for $f(r_0)$, respectively. The purple dashed line shows the large $r_0$ limit of $f(r_0) = 6R_{\lambda}/\sqrt{15}$.
(b) $[\langle{\bf r}^2(t)\rangle-{\bf r}_0^2]/f(r_0)t^2$ vs. $t/\tau_{\eta}$. The open symbols here correspond to the closed ones in (a). The red dashed line shows $(r_0^2exp[0.42(t/\tau_{\eta})]-r_0^2)/f(r_0)t^2$ with $r_0=0.9\eta$. }
\label{fig:comp2nd}
\end{center}
\end{figure}

To take a closer look at the Batchelor regime, we show in Fig.~{\ref{fig:comp2nd}}  the mean square pair separation compensated by $f(r_0)t^2$. Note that because of initial separation $r_0$ in our experiments varies continuously from the dissipative range to inertial range, we use the full function $f(r_0)= D_{LL}(r_0)+2D_{NN}(r_0)$ for the coefficient of the Batchelor scaling, instead of its dissipative range ($\frac{1}{3}r_0^2/\eta^2$) or inertial range ($\frac{11}{3}Cr_0^{2/3}/\eta^{2/3}$) scalings as in some previous studies. It is seen from the figure that  curves for all initial separations and for $t$ from $\tau_{\eta}$ to $3\tau_{\eta}$ collapse onto one horizontal line with the height very close to unity.  In the above the values of the ESFs $f(r_0)$ were independently obtained from the measured particle trajectories \cite{2011PRLNi}, which are also shown as the solid blue line in the inset of the figure. The values of $f(r_0)$ can also be obtained as the plateau heights of the compensated plots $\langle |{\bf r}(t)-{\bf r}_0|^2\rangle/t^2$ for various values of $r_0$ (not shown here), which are shown as the circles in the inset. It is seen that there is an excellent agreement between the values of the $f(r_0)$ obtained from the measured mean square pair dispersion and the Bachelor relation and those obtained directly from ESFs. Also shown in the inset are the K41 predictions for the dissipative (red line), inertial range (green line), and the large $r_0$ limit $6R_{\lambda}/\sqrt{15}$ (red dashed line) of $f(r_0)$  \cite{2008POFSawford}. The excellent collapse between circles and three solid lines indicates that the dispersion in the intermediate time domain is mainly controlled by the initial velocity difference between two particles with their separation extending from dissipative to inertial ranges. 
In Fig.~{\ref{fig:comp2nd}}(a) it is seen that the compensated $\langle |{\bf r}(t)-{\bf r}_0|^2\rangle$ for $r_{0}\sim\eta$ and $t<\tau_{\eta}$ increases systematically as $r_0$ decreases. The reason is as follows. For very small $t$ and $r_{0} \lesssim \eta$ (so the velocity difference between the pair is very small), ${\bf r}(t)$ is not much different from ${\bf r}_0$ and their difference essentially represents random measurement errors. But because of the square, these errors do not cancel but add up after averaging over different pairs. 

So far we have shown that our measured pair dispersions exhibit exponential growth in the dissipative range and power-law growth in the inertial range. However, these are manifested in different quantities, i.e. in $\langle r^2(t)\rangle$ and $\langle [{\bf r}(t)-{\bf r}_0]^2\rangle$ respectively. But in fact the pictures are consistent and there exists a crossover between the two regimes in both spatial and temporal scales.
We note that Batchelor first discussed mean square separation by using $\langle {\bf r}^2(t)\rangle-{\bf r}_0^2$ rather than $\langle |{\bf r}(t)-{\bf r}_0|^2\rangle$ \cite{1950QJRMSBatchelor}.  In Fig.~{\ref{fig:comp2nd}}(b) we plot several data sets with $r_0=0.9, 1.3, 1.7$, 2.2 and $4.3\eta$ using the original definition for pair dispersion $\langle {\bf r}^2(t)\rangle-{\bf r}_0^2$ [again normalized by $f(r_0)t^2$]. It is seen that the height of the these curves shifted downward systematically with increasing $r_0$. As $\langle r^2(t)\rangle=r_0^2exp[0.42(t/\tau_{\eta})]$ for small values of $t$ and $r_{0}$, we plot $(r_0^2exp[0.42(t/\tau_{\eta})]-r_0^2)/f(r_0)t^2$ as the dashed red line in the same figure. It is seen that even in the high-resolution compensated plot the symbols agree excellently with the line. Note that the Taylor expansion of $r_0^2exp[0.42(t/\tau_{\eta})]-r_0^2$ with respect to time is dominated by $0.42r_0^2t/\tau_{\eta}$ for $t/\tau_{\eta}<1$. This can explain why the curves for small $r_0$ tilted up in the dissipative time range for mean-square separations compensated by $t^2$. Figure~{\ref{fig:comp2nd}}(b) thus demonstrates the crossover from the exponential to the Batchelor regimes both spatially (when  $r_{0}$ varies from the dissipative to the inertial range of scales for fixed $t < \tau_{\eta}$) and temporally (when $t$ varies from smaller than $\tau_{\eta}$ to greater than $ \tau_{\eta}$ for a fixed $r_{0} < \eta$).

To summarize, we have made the first experimental study of particle pair dispersions in buoyancy-driven thermal turbulence. In the dissipative subrange of scales, our results show for the first time experimentally the existence of an exponential growth regime for the pair separation $\langle r^2(t)\rangle$, which also yield the Batchelor constant $B=0.23\pm0.07$. For time $t$ smaller and larger than $t_0$[$= (r_0^2/\epsilon)^{1/3}$], respectively, the Batchelor and the Richardson-Obukhov scalings are observed in the measured  $\langle |{\bf r}(t)-{\bf r}_0|^2\rangle$. The measured value of the Richardson constant is $g= 0.10\pm 0.07$.

We thank S.D. Huang helping the experiment and H. Xu for helpful discussions. And gratefully acknowledge support of this work by the Research Grants Council of Hong Kong under grant CUHK404409.


\begin{thebibliography}{19}
\expandafter\ifx\csname natexlab\endcsname\relax\def\natexlab#1{#1}\fi
\expandafter\ifx\csname bibnamefont\endcsname\relax
  \def\bibnamefont#1{#1}\fi
\expandafter\ifx\csname bibfnamefont\endcsname\relax
  \def\bibfnamefont#1{#1}\fi
\expandafter\ifx\csname citenamefont\endcsname\relax
  \def\citenamefont#1{#1}\fi
\expandafter\ifx\csname url\endcsname\relax
  \def\url#1{\texttt{#1}}\fi
\expandafter\ifx\csname urlprefix\endcsname\relax\def\urlprefix{URL }\fi
\providecommand{\bibinfo}[2]{#2}
\providecommand{\eprint}[2][]{\url{#2}}

\bibitem[{\citenamefont{Huber et~al.}(2001)\citenamefont{Huber, McWilliams, and
  Ghil}}]{2001JASHuber}
\bibinfo{author}{\bibfnamefont{M.}~\bibnamefont{Huber}},
  \bibinfo{author}{\bibfnamefont{J.~C.} \bibnamefont{McWilliams}},
  \bibnamefont{and} \bibinfo{author}{\bibfnamefont{M.}~\bibnamefont{Ghil}},
  \bibinfo{journal}{J. Atmos. Sci.} \textbf{\bibinfo{volume}{58}},
  \bibinfo{pages}{2377} (\bibinfo{year}{2001}).

\bibitem[{\citenamefont{Richardson}(1926)}]{1926PRSLARichardson}
\bibinfo{author}{\bibfnamefont{L.~F.} \bibnamefont{Richardson}},
  \bibinfo{journal}{Proc. R. Soc. Lond. A} \textbf{\bibinfo{volume}{110}},
  \bibinfo{pages}{709} (\bibinfo{year}{1926}).

\bibitem[{\citenamefont{Obukhov}(1941)}]{1941Obukhov}
\bibinfo{author}{\bibfnamefont{A.~M.} \bibnamefont{Obukhov}},
  \bibinfo{journal}{Izv. Akad. Nauk SSSR, Ser. Geogr. Geofiz.}
  \textbf{\bibinfo{volume}{5}}, \bibinfo{pages}{453} (\bibinfo{year}{1941}).

\bibitem[{\citenamefont{Batchelor}(1950)}]{1950QJRMSBatchelor}
\bibinfo{author}{\bibfnamefont{G.~K.} \bibnamefont{Batchelor}},
  \bibinfo{journal}{Q. J. R. Meteorol. Soc.} \textbf{\bibinfo{volume}{76}},
  \bibinfo{pages}{133} (\bibinfo{year}{1950}).

\bibitem[{\citenamefont{Sawford}(2001)}]{2001ARFMSawford}
\bibinfo{author}{\bibfnamefont{B.}~\bibnamefont{Sawford}},
  \bibinfo{journal}{Annu. Rev. Fluid Mech.} \textbf{\bibinfo{volume}{33}},
  \bibinfo{pages}{289Ð317} (\bibinfo{year}{2001}).

\bibitem[{\citenamefont{Salazar and Collins}(2001)}]{2009ARFMSalazar}
\bibinfo{author}{\bibfnamefont{J.~P. L.~C.} \bibnamefont{Salazar}}
  \bibnamefont{and} \bibinfo{author}{\bibfnamefont{L.~R.}
  \bibnamefont{Collins}}, \bibinfo{journal}{Annu. Rev. Fluid Mech.}
  \textbf{\bibinfo{volume}{41}}, \bibinfo{pages}{405} (\bibinfo{year}{2001}).

\bibitem[{\citenamefont{Batchelor}(1952)}]{1952PRSLABatchelor}
\bibinfo{author}{\bibfnamefont{G.~K.} \bibnamefont{Batchelor}},
  \bibinfo{journal}{Proc. R. Soc. Lond. A} \textbf{\bibinfo{volume}{213}},
  \bibinfo{pages}{349} (\bibinfo{year}{1952}).

\bibitem[{\citenamefont{{Ouellette} et~al.}(2006)\citenamefont{{Ouellette},
  {Xu}, {Bourgoin}, and {Bodenschatz}}}]{2006NJPOuellette}
\bibinfo{author}{\bibfnamefont{N.~T.} \bibnamefont{{Ouellette}}},
  \bibinfo{author}{\bibfnamefont{H.}~\bibnamefont{{Xu}}},
  \bibinfo{author}{\bibfnamefont{M.}~\bibnamefont{{Bourgoin}}},
  \bibnamefont{and}
  \bibinfo{author}{\bibfnamefont{E.}~\bibnamefont{{Bodenschatz}}},
  \bibinfo{journal}{New J. Phys.} \textbf{\bibinfo{volume}{8}},
  \bibinfo{pages}{102} (\bibinfo{year}{2006}).

\bibitem[{\citenamefont{Ott and Mann}(2000)}]{2000JFMOtt}
\bibinfo{author}{\bibfnamefont{S.}~\bibnamefont{Ott}} \bibnamefont{and}
  \bibinfo{author}{\bibfnamefont{J.}~\bibnamefont{Mann}}, \bibinfo{journal}{J.
  Fluid Mech.} \textbf{\bibinfo{volume}{422}}, \bibinfo{pages}{207}
  (\bibinfo{year}{2000}).

\bibitem[{\citenamefont{Ni et~al.}(2012)\citenamefont{Ni, Huang, and
  Xia}}]{2012JFMNi}
\bibinfo{author}{\bibfnamefont{R.}~\bibnamefont{Ni}},
  \bibinfo{author}{\bibfnamefont{S.~D.} \bibnamefont{Huang}}, \bibnamefont{and}
  \bibinfo{author}{\bibfnamefont{K.~Q.} \bibnamefont{Xia}},
  \bibinfo{journal}{J. Fluid Mech.} \textbf{\bibinfo{volume}{692}},
  \bibinfo{pages}{395} (\bibinfo{year}{2012}).

\bibitem[{\citenamefont{Ni et~al.}(2011)\citenamefont{Ni, Huang, and
  Xia}}]{2011PRLNi}
\bibinfo{author}{\bibfnamefont{R.}~\bibnamefont{Ni}},
  \bibinfo{author}{\bibfnamefont{S.-D.} \bibnamefont{Huang}}, \bibnamefont{and}
  \bibinfo{author}{\bibfnamefont{K.-Q.} \bibnamefont{Xia}},
  \bibinfo{journal}{Phys. Rev. Lett.} \textbf{\bibinfo{volume}{107}},
  \bibinfo{pages}{174503} (\bibinfo{year}{2011}).

\bibitem[{\citenamefont{Batchelor and Townsend}(1956)}]{1956Batchelor}
\bibinfo{author}{\bibfnamefont{G.~K.} \bibnamefont{Batchelor}}
  \bibnamefont{and} \bibinfo{author}{\bibfnamefont{A.~A.}
  \bibnamefont{Townsend}}, \bibinfo{journal}{Turbulent diffusion. In Surveys in
  Mechanics, ed. GK Batchelor, RM Davies, pp. 352-399}  (\bibinfo{year}{1956}).

\bibitem[{\citenamefont{Girimaji and Pope}(1990)}]{1990JFMGirimaji}
\bibinfo{author}{\bibfnamefont{S.~S.} \bibnamefont{Girimaji}} \bibnamefont{and}
  \bibinfo{author}{\bibfnamefont{S.~B.} \bibnamefont{Pope}},
  \bibinfo{journal}{J. Fluid Mech.} \textbf{\bibinfo{volume}{220}},
  \bibinfo{pages}{427} (\bibinfo{year}{1990}).

\bibitem[{\citenamefont{Chun et~al.}(2005)\citenamefont{Chun, Koch, Rani,
  Ahluwalia, and Collins}}]{2005JFMChun}
\bibinfo{author}{\bibfnamefont{J.}~\bibnamefont{Chun}},
  \bibinfo{author}{\bibfnamefont{D.~L.} \bibnamefont{Koch}},
  \bibinfo{author}{\bibfnamefont{S.}~\bibnamefont{Rani}},
  \bibinfo{author}{\bibfnamefont{A.}~\bibnamefont{Ahluwalia}},
  \bibnamefont{and} \bibinfo{author}{\bibfnamefont{L.~R.}
  \bibnamefont{Collins}}, \bibinfo{journal}{J. Fluid Mech.}
  \textbf{\bibinfo{volume}{536}}, \bibinfo{pages}{219} (\bibinfo{year}{2005}).

\bibitem[{\citenamefont{Boffetta and Sokolov}(2002)}]{2002PRLBoffetta}
\bibinfo{author}{\bibfnamefont{G.}~\bibnamefont{Boffetta}} \bibnamefont{and}
  \bibinfo{author}{\bibfnamefont{I.~M.} \bibnamefont{Sokolov}},
  \bibinfo{journal}{Phys. Rev. Lett.} \textbf{\bibinfo{volume}{88}},
  \bibinfo{pages}{094501} (\bibinfo{year}{2002}).

\bibitem[{\citenamefont{Yeung and Borgas}(2004)}]{2004JFMYeung}
\bibinfo{author}{\bibfnamefont{P.~K.} \bibnamefont{Yeung}} \bibnamefont{and}
  \bibinfo{author}{\bibfnamefont{M.~S.} \bibnamefont{Borgas}},
  \bibinfo{journal}{J. Fluid Mech.} \textbf{\bibinfo{volume}{503}},
  \bibinfo{pages}{93} (\bibinfo{year}{2004}).

\bibitem[{\citenamefont{Biferale et~al.}(2005)\citenamefont{Biferale, Boffetta,
  Celani, Devenish, Lanotte, and Toschi}}]{2005POFBiferale}
\bibinfo{author}{\bibfnamefont{L.}~\bibnamefont{Biferale}},
  \bibinfo{author}{\bibfnamefont{G.}~\bibnamefont{Boffetta}},
  \bibinfo{author}{\bibfnamefont{A.}~\bibnamefont{Celani}},
  \bibinfo{author}{\bibfnamefont{B.~J.} \bibnamefont{Devenish}},
  \bibinfo{author}{\bibfnamefont{A.}~\bibnamefont{Lanotte}}, \bibnamefont{and}
  \bibinfo{author}{\bibfnamefont{F.}~\bibnamefont{Toschi}},
  \bibinfo{journal}{Phys. Fluids} \textbf{\bibinfo{volume}{17}},
  \bibinfo{pages}{115101} (\bibinfo{year}{2005}).

\bibitem[{\citenamefont{Sawford et~al.}(2008)\citenamefont{Sawford, Yeung, and
  Hackl}}]{2008POFSawford}
\bibinfo{author}{\bibfnamefont{B.~L.} \bibnamefont{Sawford}},
  \bibinfo{author}{\bibfnamefont{P.~K.} \bibnamefont{Yeung}}, \bibnamefont{and}
  \bibinfo{author}{\bibfnamefont{J.~F.} \bibnamefont{Hackl}},
  \bibinfo{journal}{Phy. Fluids} \textbf{\bibinfo{volume}{20}},
  \bibinfo{pages}{065111} (\bibinfo{year}{2008}).

\bibitem[{\citenamefont{Schumacher}(2008)}]{2008PRLSchumacher}
\bibinfo{author}{\bibfnamefont{J.}~\bibnamefont{Schumacher}},
  \bibinfo{journal}{Phys. Rev. Lett.} \textbf{\bibinfo{volume}{100}},
  \bibinfo{pages}{134502} (\bibinfo{year}{2008}).

\end{thebibliography}

\end{document}